# EXPLORING IOT FOR REAL-TIME $CO_2$ MONITORING AND ANALYSIS

*Authors (A-Z): Abhiroop Sarkar, Debayan Ghosh, Kinshuk Ganguly, Snehal Ghosh, Subhajit Saha*

*Affiliation: Department of Computer Science and Engineering (Artificial Intelligence and Machine Learning), Institute of Engineering & Management, Salt Lake, Kolkata-700091*

**Abstract.** As a part of this project, we have developed an IoT-based instrument utilizing the NODE MCU-ESP8266 module, MQ135 gas sensor, and DHT-11 sensor for measuring $CO_2$ levels in parts per million (ppm), temperature, and humidity. The escalating $CO_2$ levels worldwide necessitate constant monitoring and analysis to comprehend the implications for human health, safety, energy efficiency, and environmental well-being. Thus, an efficient and cost-effective solution is imperative to measure and transmit data for statistical analysis and storage. The instrument offers real-time monitoring, enabling a comprehensive understanding of indoor environmental conditions. By providing valuable insights, it facilitates the implementation of measures to ensure health and safety, optimize energy efficiency, and promote effective environmental monitoring. This scientific endeavor aims to contribute to the growing body of knowledge surrounding $CO_2$ levels, temperature, and humidity, fostering sustainable practices and informed decision-making.

## 1 Introduction

To ensure sound health, appreciable productivity, energy efficiency, and ecological sustainability it is essential to monitor and analyze the levels of $CO_2$ in an indoor environment. As a result of anthropogenic activities, rising $CO_2$ levels have become a pertinent concern for us. Owing to its adverse effects on health for instance fatigue, respiratory ailments, and reduced cognitive function. An increase in the level of $CO_2$ emissions leads to climate change making the development of efficient and cost-effective models indispensable.

Through this paper, we intend to present the development of an IoT-based instrument for measuring $CO_2$ levels in ppm, temperature, and humidity. The devices utilized for the system include the NODE MCU-ESP8266 module, MQ135 gas sensor, and DHT-11 sensor for real-time data collection and analysis.

The instrument continuously measures $CO_2$ levels and provides real-time data for analysis that can both serve for later analysis and also facilitate prompt steps to restore air quality conditions to normal.

To ensure accurate $CO_2$ measurements, the MQ135 gas sensor is employed, which exhibits high sensitivity to $CO_2$ and provides reliable results. The DHT-11 sensor is utilized to monitor the temperature and humidity. The implementation of IoT (Internet of Things) technology enables the seamless collection and transmission of data to the ThingSpeak portal for long-term preservation and statistical analysis. This leads to comprehensive monitoring and evaluation of indoor environmental conditions. We aim to contribute to the scientific understanding of $CO_2$ levels, temperature, and humidity monitoring in indoor environments by providing valuable insights, the developed instrument offers opportunities for informed decision-making regarding health and safety measures, energy efficiency improvements, and environmental conservation strategies.

By providing valuable insights, the developed instrument offers opportunities for informed decision-making regarding health and safety measures, energy efficiency improvements, and environmental conservation strategies.

In this paper we provide detailed accounts of the materials and methods employed, the circuit setup, and the working results alongside specific test cases. The results obtained can serve as significant resource for researchers, policymakers and professionals of different domains and allow them to implement better safety standards and procedures to control the rising $CO_2$ emission levels.

## 2 Components Used

- NodeMCU ESP 8266 Wi-Fi model and microcontroller
- DHT 11 Temperature and Humidity Sensor
- MQ 135 gas sensor
- LCD (16x2) display I2C module
- Jumper Cables
- Led
- Breadboard
- Plastic Box to safely enclose the contents
- Thing Speak

## 3 Objectives

1. Develop an IoT-based instrument for measuring $CO_2$ levels, temperature, and humidity.

2. Continuously monitor and provide real-time data on indoor environmental conditions.

3. Detect and alert high $CO_2$ levels to improve health and safety.

4. Improve comfort and productivity in indoor environments.

5. Enable remote monitoring and adjustment of environmental conditions.

6. Incorporate an LCD display and buzzer for ease of use and alerts.

## 4 Procedure

1. The code was written in Arduino IDE with specifications appropriate for the ESP8266 module and uploaded to the in-built microcontroller on the board.

2. The MQ135 Gas Sensor was placed in a room with fresh air and proper circulation for 15 hours to enable proper calibration. The Ro and Rs values were then adjusted in the library code.

3. All the sensors are connected to NodeMCU by attaching their positive supply pin to Vcc or 3V3 pin of MCU and negative to GND (ground) pin of NodeMCU.
4. The Gas sensor's analog pin was connected to Ao pin of NodeMCU, while the power and ground pins were connected to the Ground and 3V supply pins of NodeMCU.
5. The DHT sensor was connected in the same manner as the gas sensor, with the exception of the output pin being connected to Digital Pin D3 of ESP8266.
6. The LCD I2C Module's power and ground pins were connected to the Vu and Ground pins of the Node MCU.
7. The SDA and SCL pins were connected to Digital pins D1 and D2 of the NodeMCU.
8. A buzzer and a red LED were connected to the ground and D8 pin of the Node MCU, and a green LED was connected to D7, serving as the alarm system.
9. All connections were checked, and the entire setup was securely placed in a box made of hard plastic.
10. The ThingSpeak API, Channel ID, Wi-Fi SSID are other details are carefully incorporated in the code.
11. The functionality and the data on ThingSpeak are monitored regularly to identify any loopholes or scopes of improvement.

## 5 Circuit design

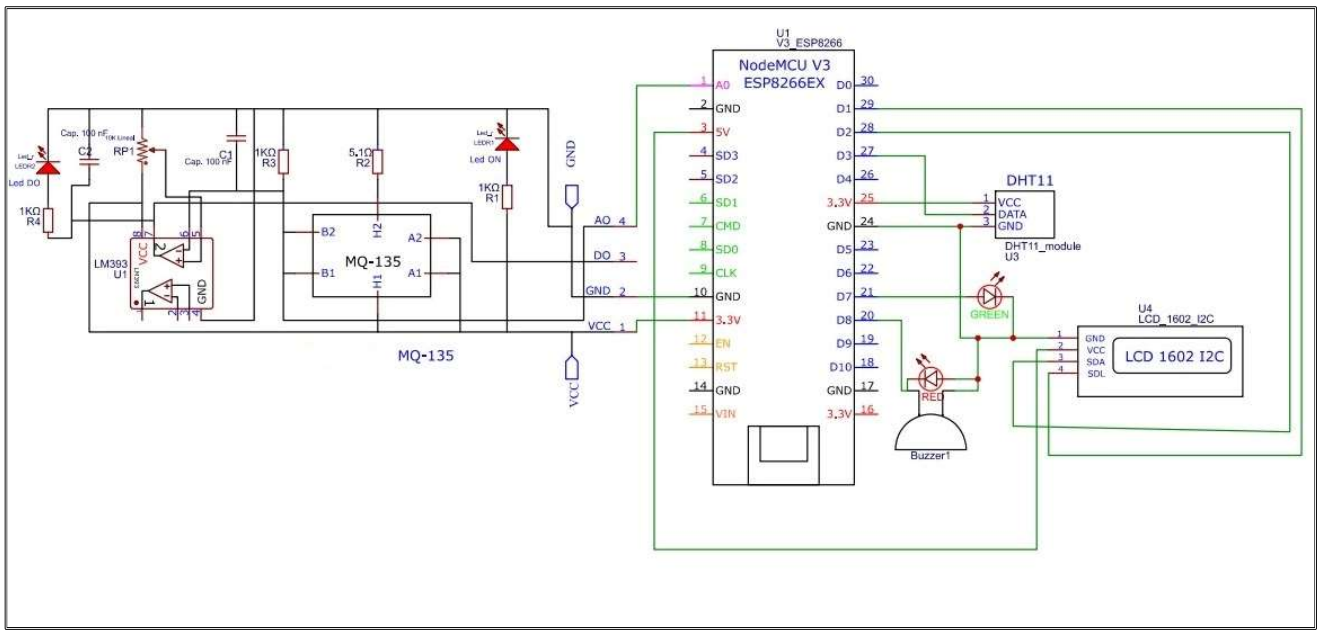

Circuit Diagram

## 6 Code

```
#include <MQ135.h>
#include <LiquidCrystal_I2C.h>
#include <MQ135.h>
#include<Wire.h>
#include <SoftwareSerial.h>
#include "ESP8266WiFi.h"
#include <WiFiClient.h>
#include <ThingSpeak.h>
#include "DHT.h"        // including the library of DHT11 temperature and humidity sensor
#define DHTTYPE DHT11   // DHT 11
#include <LiquidCrystal_I2C.h>
LiquidCrystal_I2C lcd(0x27,16,2);
#define dht_dpin 0
#define buz 15
#define lg 13
const char* ssid = "WiFi SSID";      // WiFi SSID
const char* password = "wifi password";  // WiFi password
const char* server = "api.thingspeak.com"; // ThingSpeak server
const char* apiKey = "4TRP63OXJX7B45Y6";        // ThingSpeak API key
WiFiClient client;

DHT dht(dht_dpin, DHTTYPE);
void setup(void)
{ pinMode(buz, OUTPUT);
  pinMode(lg, OUTPUT);
  dht.begin();
  Serial.begin(9600);
  ThingSpeak.begin(client);
  WiFi.begin(ssid, password);
  digitalWrite(lg,HIGH);
  while (WiFi.status() != WL_CONNECTED)
  {
    delay(1000);
    Serial.println("Connecting to WiFi...");
  }
  Serial.println("Connected to WiFi.");
  lcd.begin(16,2);
  lcd.init();
  lcd.backlight();
}
void loop() {
```

```
  `
MQ135 gasSensor = MQ135(A0);
int h = dht.readHumidity();
float t = dht.readTemperature();
int ppm = analogRead(A0);
float R1=gasSensor.getRZero();
float R2=gasSensor.getResistance();
float CO2ppm=gasSensor.getCorrectedPPM(t, h);

Serial.print("Current humidity = ");
Serial.print(h);
Serial.print("%  ");
Serial.print("temperature = ");
Serial.print(t);
Serial.print("°C  ");
Serial.print(" analog sensor value = ");
Serial.print(ppm);
Serial.print("Ro=");
Serial.print(R1);
Serial.print("Rs=");
Serial.print(R2);
Serial.print(" CO2 PPM= ");
Serial.println(CO2ppm);
lcd.clear();

lcd.setCursor(0,0);
lcd.print("Temp=");
lcd.print(t);
lcd.setCursor(10,0);
lcd.print("C");
lcd.setCursor(0,1);
lcd.print("Humidity=");
lcd.print(h);
lcd.setCursor(12,1);
lcd.print("%");
delay(2000);
lcd.clear();
lcd.setCursor(0,0);

if(CO2ppm>500)
{
 lcd.print("CO2 PPM-");
 lcd.print(CO2ppm);
 lcd.setCursor(0,1);
 lcd.print("HIGH PPM ALERT");
```

```
 digitalWrite(buz,HIGH);
 digitalWrite(lg,LOW);
 }
 else
 { digitalWrite(lg,HIGH);
  digitalWrite(buz,LOW);
lcd.print("CO2 PPM-");
lcd.print(CO2ppm);
 }
 delay(2000);
 lcd.clear();
 ThingSpeak.setField(1, t);
 ThingSpeak.setField(2, h);
 ThingSpeak.setField(3, CO2ppm);
 int status = ThingSpeak.writeFields(2097285, apiKey);
 if (status == 200)
 {
   Serial.println("Updated ThingSpeak channel successfully.");
 }
 else
 {
   Serial.println("Failed to update ThingSpeak channel.");
 }
 }
```

## 7  Results

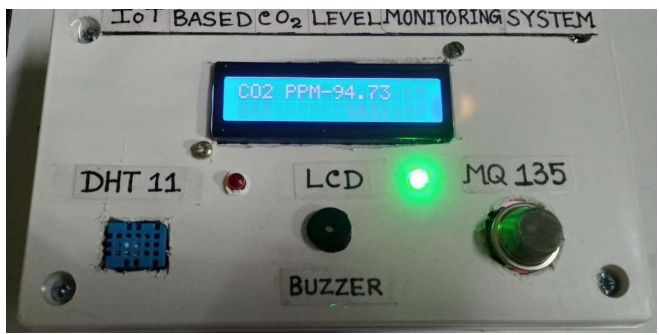

Normal $CO_2$ level

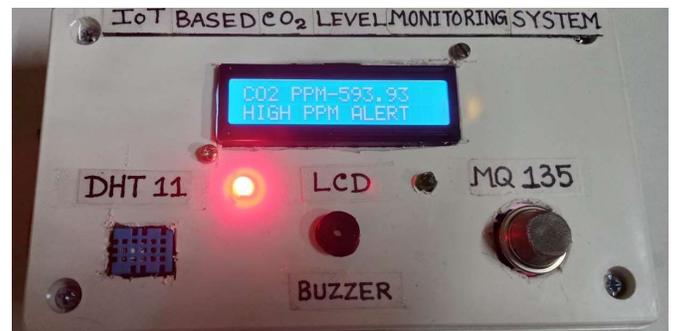

High $CO_2$ level

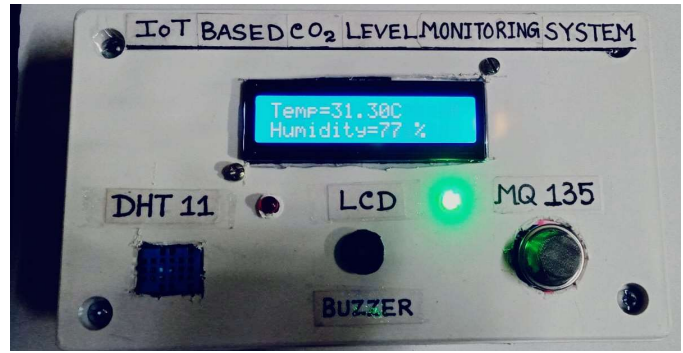

Temperature and Humidity Level

Real-Time Data Plotting using ThingSpeak (Temperature-Time Plot):

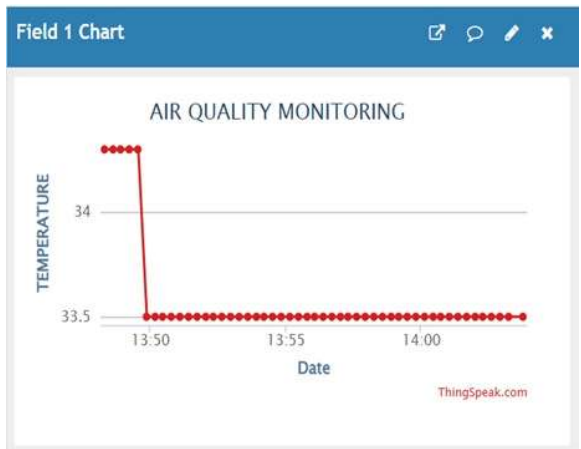

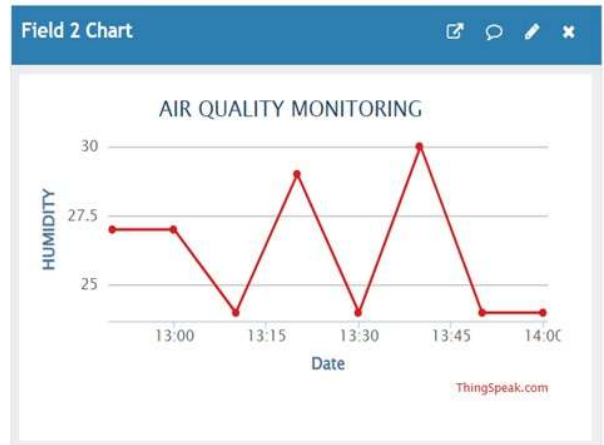

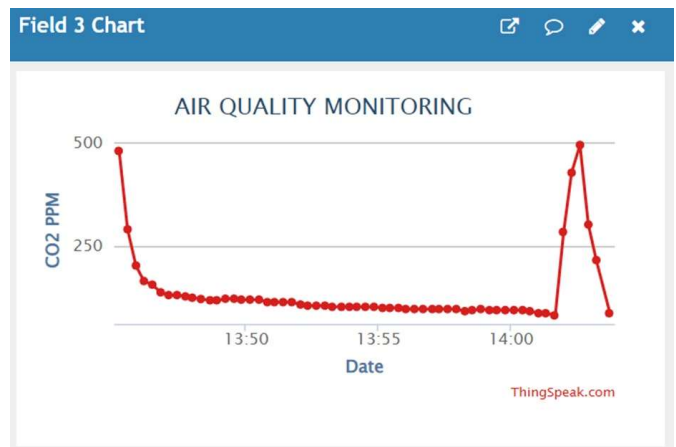

## 8  Probable places of use

An IoT-based instrument for measuring $CO_2$ levels, temperature, and humidity is of significant importance in various fields. Here are some key points that highlight the importance of such an instrument:

1. Health and Safety: High levels of $CO_2$ in indoor environments can have adverse effects on human health, such as headaches, fatigue, and reduced cognitive function. An IoT-based instrument can continuously monitor $CO_2$ levels and provide alerts when they exceed safe levels, thus improving the safety and health of individuals.
2. Energy Efficiency: An IoT-based instrument can measure temperature and humidity levels and provide insights into the efficiency of HVAC systems. By identifying areas of inefficiency, building managers can make targeted adjustments to reduce energy consumption and costs.
3. Environmental Monitoring: $CO_2$ is a greenhouse gas that contributes to climate change. By continuously monitoring $CO_2$ levels, IoT-based instruments can provide valuable data for policymakers and researchers to develop effective climate change mitigation strategies.
4. Quality Control: In certain industries such as food and pharmaceuticals, temperature and humidity levels must be carefully controlled to maintain product quality. An IoT-based instrument can ensure that these conditions are met, reducing waste and ensuring product consistency.
5. Comfort and Productivity: Optimal temperature and humidity levels can improve comfort and productivity in indoor environments. An IoT-based instrument can provide real-time data to building managers to ensure that these conditions are maintained.
6. Remote Monitoring: An IoT-based instrument can be accessed remotely, providing valuable data for building managers or researchers who may be located in a different location. This feature allows for easy monitoring and adjustment of environmental conditions, improving efficiency and accuracy.

## 9  Conclusion

In conclusion, the development of an IoT-based instrument for measuring $CO_2$ levels, temperature, and humidity has significant implications for indoor environmental monitoring. By continuously monitoring and providing real-time data on these parameters, the instrument can improve health and safety, energy efficiency, environmental monitoring, and product quality. The incorporation of an LCD display and buzzer provides ease of use and alerts, making it an invaluable tool for various applications. Furthermore, the use of Arduino and NodeMCU-ESP866 microcontrollers along with MQ135 and DHT11 sensors demonstrates the versatility and adaptability of IoT-based technology for environmental monitoring.

## 11  Acknowledgement